\documentclass{article}

\usepackage{graphicx}
\usepackage{authblk}
\usepackage{latexsym}
\usepackage{amssymb}
\usepackage{bm} 
\usepackage{url}

\begin{document}

\title{Spectral analysis for gene communities \\ in cancer cells}


\author{{
\sc Ayumi Kikkawa}$^*$,\\
Mathematical and Theoretical Physics Unit, \\
Okinawa Institute of Science and Technology Graduate University,\\
1919-1 Tancha, Onna-son, Kunigami-gun, Okinawa, 904-0495 Japan \\
$^*${\small{Corresponding author: akikkawa@oist.jp}}
}

\maketitle

\begin{abstract}
{We investigate gene interaction networks in various cancer cells by spectral analysis of the adjacency matrices.  We observe localization of the networks on hub genes which have extraordinarily many links.  The eigenvector centralities take finite values only on special nodes when the hub degree exceeds a critical value $d_c \simeq 40$.  The degree correlation function shows the disassortative behavior in the large degrees, and the nodes whose degrees $d \gtrsim 40$ have tendencies to link to small degree nodes. 
The communities of the gene networks centered at the hub genes are extracted by the amount of node degree discrepancies between linked nodes.  We verify the Wigner-Dyson distribution of the nearest neighbor eigenvalues spacing distribution $P(s)$ in the small degree discrepancy communities, and the Poisson $P(s)$ in the communities of large degree discrepancies including the hubs.
}
{graph spectra; vector centrality; adjacency matrix; community structure; random matrix model}
\\
\end{abstract}

\section{\label{sec:01}Introduction}

In recent decades, many works have been done to probe network structures in large fields including WWW or internet, social networks and biological networks.  Among them, the investigations of spectra of various matrices expressing graphical topologies of the networks have been under enthusiastic studies.  A graph which represents a network consists of nodes (or vertices) and links (or edges).  The adjacency matrix of size $N$, where $N$ is the number of nodes in the graph, is constructed by setting its $(i,j)$ element as $A_{ij}=1$ if there is a link between the nodes $i$ and $j$, or $A_{ij}=0$ otherwise.  In the case of directed networks, the links are replaced by arrows.  In this paper, we discuss the undirected networks, the adjacency matrix $A$ becomes a real symmetric matrix.  All eigenvalues $\lambda_i$ are real and in the large-$N$ limit, their distribution $\rho(\lambda)$ shows the famous Wigner's semi-circle independent of the graph details  \cite{Mehta1991, Guhr1998}.
\begin{equation}
\rho(\lambda)\propto \sqrt{2N-\lambda^2 }
\end{equation}
This universal behavior in the large-$N$ limit obtained from the random matrix theory, requires that the ratio $L/N$ to be constant, where $L$ is the number of links, so the networks should become  considerably dense in this limit.  

In contrast to this universality, the spectra of adjacency matrices or other graph representative matrices of the real-world networks  are somewhat different.  Since interaction resources are restricted, the number of links should be limited.  Therefore, the large-$N$ limit of the complex real-world networks is expected to be a very sparse network.  On the other hand, the spectra of the sparse random matrix have been investigated, and the deviation from the Wigner's semi-circle law has been discussed \cite{Kuhn2008, Mirlin1991, Rodgers1988, Nagao2008}.  The  center of the spectral band rises sharper and long tails appear in both edges.  In the scale-free networks, the power-law (long tail) behavior in the edge of the spectra is given by \cite{Dorogovtsev2003}
\begin{equation}
\label{Eq:powers}
\rho(\lambda)\propto \lambda^{-\mu} , \qquad \mu = 2\gamma -1
\end{equation}
where $\gamma$ is the power of the distribution $P(k)$ of node degrees $k$ in corresponding graph,
\begin{equation}
P(k)\propto k^{-\gamma} .
\end{equation}
This power-law distribution $P(k)$ is also well-known feature of the scale-free networks \cite{Barabasi2002, Barabasi2016}. It is conjectured that many real-world networks including the biological networks share such scale-free features.  In spite of many studies with numerically simulated model networks or real-world networks obtained from huge empirical data, the theoretical basis of this universality is still unclear.   

The largest eigenvalue of the adjacency matrix $\lambda_\mathrm{max}$ has a relation to the node which has the largest degree $d\equiv k_\mathrm{max}$ so called a hub, and sometimes forms a distinguishable narrow band above the bulk of the spectrum. In this case, the network is 'localized' around the hub, in relation to impurity bands studied in the condensed matter physics.  Furthermore, several peaks at the band center are formed by many zero eigenvalues of star-graphs, in which one node of degree $k$ is connected to other $k=1$ nodes \cite{Bauer2001}. These peaks have close relationships to the modular structure of the network.  Here we use the term ‘modular’ to mention community structures of the network.   In a network community, the nodes are bounded together more frequently inward compared to outward.  Also, the eigenvector corresponding to the second largest eigenvalue is called the largest diffusive modes and have been investigated with the method of a random walk model on a lattice \cite{Eriksen2003, Simonsen2004}.

Kinds of measures, such as degree centrality, eigenvector centrality, modularity, bipartivity and assortativity, have been used to investigate the community structures of complex networks so far. For reviews see Refs. \cite{Newman2004, Estrada2010}.  In the case of the gene interaction network which consists of more than 20,000 nodes and several hundreds of thousands edges,  it is inevitable to divide the network into sub-networks and extract some community structures including some specific nodes.  Extraction of some gene communities in cancer cells might be extremely useful for detection of gene groups that determine cell characteristics, classification of pathological conditions, selection of therapeutic methods, and so on.  Also it has been pointed out that the marginal genes which connect different communities owe some important role in the cell \cite{Newman2006}.  

Gene network data associated with cancer cells are available from several public databases. One of them, the TCNG database (http://tcng.hgc.jp) stores gene co-expression networks numerically inferred by Bayesian network algorithms called SiGN-BN NNSR \cite{Tamada2011}. Using these data, we have found the spacing distribution of two succeeding eigenvalues $P(s)$ of the adjacency matrices follows the Wigner distribution \cite{Kikkawa2018}. This is the result in a dense gene network group, where the number of inferred edges is large. 

In this work, the same adjacency matrices are used to examine the community structure of the gene networks. In particular, when the degree of the hub node becomes larger than a certain threshold $d > d_c$, the network localization on the hub nodes is observed. Furthermore, we evaluate the node degree correlation function. We show that some disassortative gene communities localized around the hub can be extracted. 

\section{Methods}
\subsection{Gene interaction networks in TCNG database}
In this study, we use  numerically inferred gene interaction networks each of which consists of 8,000 nodes downloaded from the TCNG database. These gene networks were obtained from huge calculations based on the Bayesian network model. The original expression data of human cancer cell samples used for the gene network inference were taken from the GEO database \cite{Barrett2012}.  The number of expression samples used for a computational assay of the Bayesian network inference is around 100. The samples are taken mainly from various types of cancer (tumor) cells including a small portion of normal cells as controls. In the Bayesian networks, the interactions (or causal dependencies) between the nodes are represented by arrows, and the direction corresponds to the causality of the interaction. In this study we ignore the direction of gene interactions (edges) for simplicity. We also ignore self-interactions. Then the adjacency matrix $A$ of the gene interaction network becomes a real symmetric matrix.

The degree $k_i$ of node $i$ is given by the sum of each row (or column) of the adjacency matrix $A$.
\begin{equation}
k_i =\sum_j A_{ij}    
\end{equation} 
The mean degree $\bar{k}$ of all gene expression networks used in this study is $\bar{k} =9.68$, which is within the range where the sparse random matrix predictions apply.

\subsection{Eigenvector centrality and the inverse participation ratio}
The eigenvector centrality $v_i$ is one of the measures often studied in a framework of the graph theory to probe network properties \cite{Nadakuditi2013, Martin2014}. It is evaluated  from the eigenvector corresponding to the largest eigenvalue $\lambda_1$ by diagonalizing the adjacency matrix $A$.
\begin{equation}
AV = {\bf \lambda}V,  \quad V=[v^1, v^2, \cdots, v^N ], \quad {\bf \lambda}=(\lambda_1 , \cdots , \lambda_N ),
\end{equation}
where we label the eigenvalues $\lambda_i$ in descending order $\{\lambda_1 > \lambda_2 > \cdots > \lambda_N\} $  and $v^i $ $(i=1\cdots N)$ is the corresponding eigenvectors.  The centrality of node $i$ is given by the $i$-th element of $v^1$.
\begin{equation}
v_i \equiv v^1_i
\end{equation}

Its variant is also known as the Google's PageRank.  
In the localized networks where the hub degree $d$ becomes very large, the eigenvector centralities take finite values only on the hub node and the nodes connected to the hub. 
It takes the values of $O(1/N)$ on other nodes.  
Each of our gene networks has a different mean degree. In the original database TCNG, eacn edge has an edge attribute named 'edge factor', which corresponds to the likelihood of the inference. To fix the mean degree $\bar{k}$ of all networks, the edges are extracted in the order from larger to smaller edge factors.

We also evaluate the inverse participation ratio $\Psi$ which is obtained by
\begin{equation}
\Psi (\lambda_i ) = \sum_j (v^i_j )^4 .
\end{equation}
We note that all eigenvalues are normalized as $|v|^2 = 1$.
In the localized regime, it takes finite values $\Psi \sim O(1)$ only on localized elements of $\lambda_i$.   

\subsection{The degree correlation fuction}
By averaging several gene interaction networks obtained from the gene expression experiments in various cancers, it is able to see a coarse-grained property of the gene networks in cancer cells.  Let $ h $ be the maximum node degree of all 254 gene networks, and we evaluate the degree correlation matrix $ E $ of size $ h \times h $. Here, the element $ E_ {ij} $ is  the probability that two nodes have degrees $k=i$ and $k=j$, respectively, at each end of an edge taken randomly from the whole network.  
\begin{equation}
 \sum_{i,j=1}^h E_ {ij} = 1 
 \end{equation}
The sum is taken over all edges in 254 gene interaction networks.  The degree correlation function $k_{nn}(k) $ is evaluated from the correlation matrix $E$ by
\begin{eqnarray}
k_{nn} (k) &=&\sum_{k'} P(k' |k) \\
P(k' | k) &=& \frac{E_{kk'}}{\sum_{k'}E_{kk'}}.
\end{eqnarray}
\cite{Pastor-Satorras2001, Costa2007, Newman2002}
When the degree correlation function $k_{nn}(k)$ is plotted against the degree $k$,  the positive slopes have been obtained in the case of social networks including the collaborative research networks. They are called the degree assortative networks.  On the contrary, in the metabolic network of E. coli or in some other biological networks, the degree dissasortative behavior is obtained.  In this case, the hub which has large degree has tendency to connect to small degree nodes, and $k_{nn}(k)$ shows the negative slope in the log-log plot.  The neutral degree correlation networks are also known, for example, in a power grid network \cite{Newman2003, Jalan2015}.

$ k_ {nn}(k) $ is also expected to follow a power-law, which has been observed in many studies,
\begin{equation}
 k_ {nn} (k) \propto\  k ^{\eta}.
\end{equation}
 We evaluate the power $ \eta $ from the average over all edges of the whole gene network in the 254 cancer studies.

\subsection{Nearest neighbor eigenvalues spacing distribution $P(s)$}
In the random matrix theory, in addition to the eigenvalue density, the universal distribution of nearest level spacing $P(s)$ is important \cite{Akemann2011a}. In the case of the real symmetric random matrices, the Wigner distribution is 
\begin{equation}\label{wd}
P(s) = \frac{\pi s}{2}\exp \left( -\frac{\pi s^2}{4} \right).
\end{equation}
This universal behavior has been widely observed in the adjacency matrices of real-world networks including the biological networks .   Eq.(\ref{wd}) also tells that there is  correlation between the eigenvalues.  On the other hand, when the eigenvalues distribute randomly with no correlation,  for example, if the eigenvalues are taken from the Poisson distributed random values, their interval distribution $P(s)$ becomes
\begin{equation}\label{poisson}
P(s) = \exp (-s),
\end{equation}
where $s$ is the spacing between two adjacent eigenvalues normalized by the mean eigenvalue spacing.  Eq.(\ref{poisson}) is called the Poisson distribution in the random matrix theory.

Before evaluating the intervals between two succeeding eigenvalues of the adjacency matrices,  we 'unfold' the original eigenvalues numerically.  We divide eigenvalues $\lambda_1 > \lambda_2 > \cdots > \lambda_N $ into segments each of which consists of several hundreds of eigenvalues. Afterward, we rescale $\lambda_i$ according to the local mean spacing $\bar{s}$ in each segment.  With these unfolded eigenvalues, local distributions of nearest neighbor level spacing are obtained.  The Kolmogorov-Smirnov test for these local distributions against the null-distributions eqs.(\ref{wd}) and (\ref{poisson}) has been performed with the significance level $\alpha =0.05$.  For the details of the unfolding method, see ref.\cite{Kikkawa2018}.

\section{Results}
\begin{figure}[!h]
\centering
\includegraphics[width=3.5in]{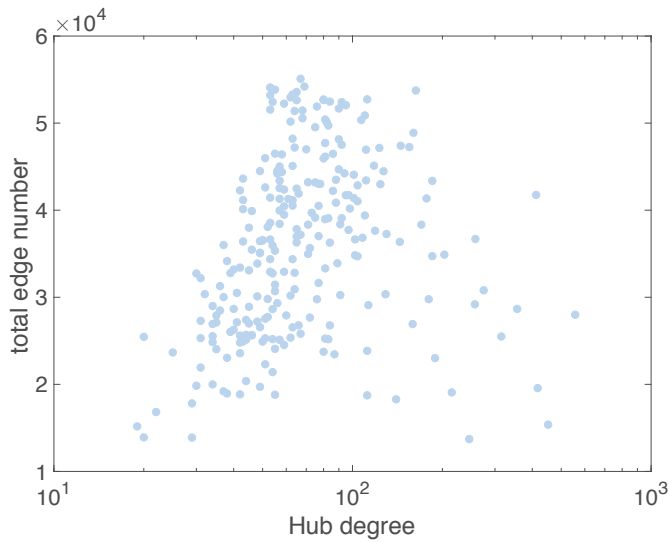}
\caption{For all 254 gene networks, the number of edges are plotted against the degree of the hub node $d$.  The number of nodes is fixed as $N\simeq 8000$ for all gene networks. }
\label{Fig:01}
\end{figure}

Figure \ref{Fig:01} shows the relation between the total edge number in the inferred 254 gene interaction networks and the hub degree $d$.  Up to $d \lesssim 100$, the hub degree grows almost linear to the total number of edges in the network.  However in the region where $d \gtrsim 100$,  the hub degrees are almost independent of the total edge numbers, and the average node degree $\bar{k}$ of the networks becomes smaller $\bar{k}=8.4$ (where  $\bar{k}=9.7$ for the whole network).  In the gene networks data, the number of nodes $N$ is fixed as $8000$.  It can be predicted that the edges concentrate more and more on the hub nodes in much sparser gene interaction networks where $N$ become much larger.

\begin{figure}[!h]
\centering
\includegraphics[width=3.5in]{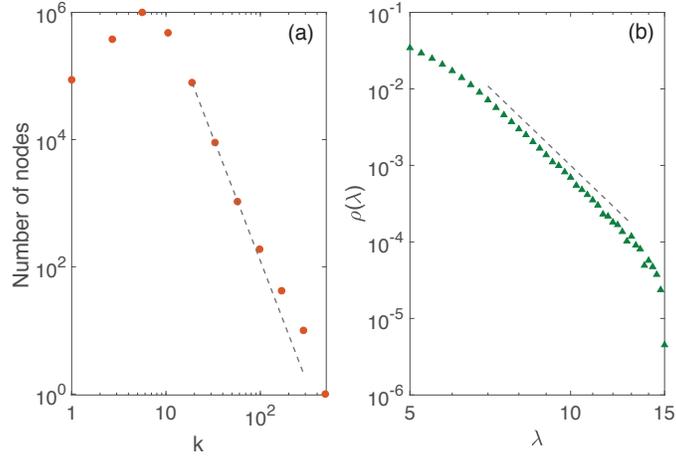}
\caption{{\rm (a)} The distribution of node degree $P(k)$ in the log-log scale.  The logarithmic bins are used and the $n$-th bin edge is $k^{(n)}= (1.7^n -1)/0.7$.  In this figure $1 \le n \le 11$ bins are shown.  The curve fitting by a power function for $5$th to $10$th bins are performed. The result $P(k)\propto  k^{-3.8}$ is plotted by the dotted line.  
{\rm (b)} The eigenvalue density $\rho(\lambda)$ in the tail region $\lambda > 5$. The bin width is $0.25$.  The power-law behavior $\rho(\lambda)\propto \lambda^{-6.7}$ is also shown by the dotted line as a guide to the eye.  }
\label{Fig:02}
\end{figure}

In FIG.\ref{Fig:02}(a) and (b), we plot the degree distribution $P(k)$ and the spectrum $\rho(\lambda)$ in log-log scale,  where $k$ is the node degree and $\lambda$ is the eigenvalue of adjacency matrix $A_{ij}$, respectively.  To obtain the histogram of $k$ in FIG.\ref{Fig:02}(a), the logarithmic bins are used. The number of corresponding nodes decreases rapidly in larger bins, so the width of $n$-th bin is taken proportional to $c^n$ where $c$ is some number $c>1$.
The $n$-th bin edge is $k^{(n)}=(c^n -1)/(c-1)$.  Using the MATLAB function 'fit' \cite{MATLAB2018a}, we evaluate the power $\gamma$ for several values of $c$ and over some succeeding bins.  The minimum width of $95\%$ confidence interval of fitted $\gamma$ is obtained when $c=1.7$ and with the data of $n=5$ to $10$ bins, which gives the result $\gamma=3.87 \pm 0.01$.

In FIG.\ref{Fig:02}(b), we plot the tail part of $\rho(\lambda)$ for $\lambda\ge 5$.  Here we take a constant bin width $0.25$.  We apply the same fitting procedure for the power $\mu$.  The minimum of the $95\%$ confidence interval is obtained in the region $7 < \lambda < 13$, which gives the power $\mu = 6.52 \pm 0.05$.  
Although our fitting procedure may not precise enough, the value of $\mu$ is just slightly smaller than the expected from the relation in eq. (\ref{Eq:powers}).

In Ref.\cite{Nadakuditi2013, Martin2014},  the eigenvalue centrality $v_i$ is investigated for a model network where the node degree $k$ is taken as Poisson distributed random values, the Poisson random graph. In the random graphs, the eigenvalue centralities $v_i$ are $O(1/\sqrt{N})$ for all $i=1,\cdots, N$.  However, by adding an extra hub node of degree $d$, it was shown analytically that $v_i$ takes $O(1)$ value only on the hub and its neighbor nodes, and  $v_i \sim O(1/N)$ on any other nodes.  The square of the eigenvector centrality in the large $N$ limit is given by,
\begin{equation}
  v_{i}^2  = \left\{ \begin{array}{ll}
    (\frac{1}{2}d-c)/(d-c) & \mbox{for $i=1$ (the hub),} \\
    (\frac{1}{2}d-c)/(d-c)^2 & \mbox{for $i$ a neighbor of the hub,} \\
     0 &  \mbox{otherwise}
  \end{array} \right.
  \label{Eq:vi2}
\end{equation}
where $c$ is the mean degree $\bar{k}$ and $d>2c$.  Note that the eigenvalues are enumerated in descending order.   

\begin{figure}[!h]
\centering
\includegraphics[width=3.5in]{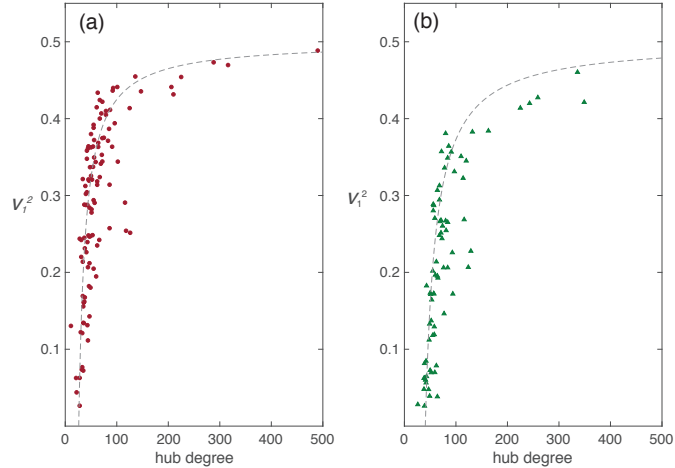}
\caption{{\rm (a)} $v_1^2$ v.s. the hub degrees are plotted for 112 gene interaction networks where the mean degree $c=4.7$ . The dashed line is the result of a curve fitting by the function in eq.(\ref{Eq:vi2}). The fitting parameter is $c^\dagger = 13$.  The localization transition at $d_c =2c^\dagger = 26$ is larger than the theoretical prediction $d_c = 9.4$.   
{\rm (b)} The same plot for the networks where $c=6.6$ and we obtain $d_c = 40$ from the fitting (the dashed line). The number of networks plotted in this figure is 76.}
\label{Fig:03}
\end{figure}

To see this localization properties, we plot $v_1^2$ versus the hub degrees $d$ of the gene interaction networks.  In FIG.\ref{Fig:03}, two results of different $c$ ($=\bar{k}$) values $c=4.7$ and $c=6.6$ are shown.  Before diagonalizing the adjacency matrix $A$,  the edges are extracted according to the edge factors from high to low values, thus more confidential edges in the numerical inference are preferred, until the average degree attains the fixed values.  We employ the curve fitting for the results by eq.(\ref{Eq:vi2}) of $i=1$ case, setting $c$ as a fitting parameter.  As seen from the dotted lines in FIGs.\ref{Fig:03}(a) and (b), the fitting parameters $c^\dagger$ become larger than empirical value of $c$ in both cases although the total behavior is well described by the theoretical result.  The localization transition is expected at $d_c =26$ and $40$ from the fitting, respectively, where $d_c = 2 c^\dagger$.  The data points shown in the figures are evaluated with the gene interaction networks which have edges more than $L > {c N}/{2}$.

Since the gene networks are the scale-free networks, which is shown by the power-law distribution $P(k)$ in FIG.\ref{Fig:02}(a), there are correlations between node degrees.  We note that in the study of the node disassortativity of the scale-free networks \cite{Newman2002},  transition to one highly connected graph (the giant component formation) at the value of $\bar{k}=1$ also shifts to a larger value of $\bar{k}>1$. The nodes of larger degrees tend to connect to lower degree nodes, thus prevent the network to form one fully connected graph.  The larger values of $d_c$ for the expected localization transition of the gene interaction networks might be also explained by this dissasortative scale-free network behavior.  
We also note that in the dense gene interaction networks of higher mean degree $c > 8$,  the discrepancy from eq.(\ref{Eq:vi2}) becomes very large and the curve fitting method does not work anymore.  In such dense gene networks,  the correlation of node degrees is very large and the assumption of uncorrelated edge degree of the Poisson random graph for the derivation of eq.(\ref{Eq:vi2}) cannot be applied.

\begin{figure}[!h]
\centering
\includegraphics[width=3.5in]{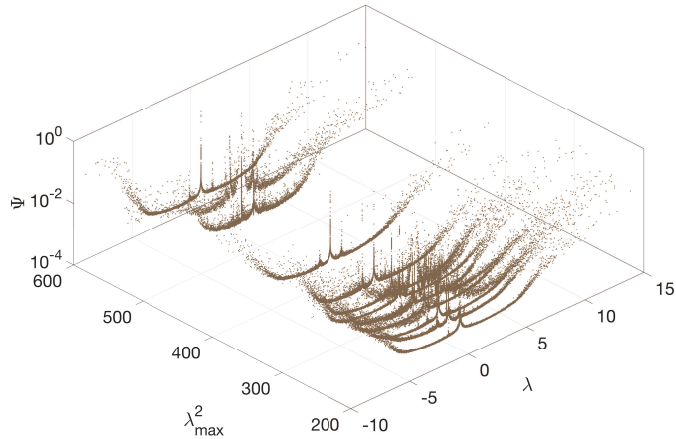}
\caption{The scatter plot of the inverse participation ratio $\Psi(\lambda_i )$ for the strongly localized gene interaction networks which have the largest eigenvalue $\lambda_\mathrm{max} ^2 >200$.}
\label{Fig:04}
\end{figure}

The inverse participation ratio $\Psi(\lambda_i )$ is plotted in log-scale in FIG.\ref{Fig:04}  for the gene networks that have $\lambda_\mathrm{max}^2 > 200$, where $\lambda_\mathrm{max}\equiv \lambda_1$ is the largest eigenvalue of $A$.  It is also given by 
\begin{equation}
\lambda_\mathrm{max} \simeq\sqrt{d}
\end{equation}
where $d$ is the hub degree \cite{Nadakuditi2013}.  The scatter plot in FIG.\ref{Fig:04} shows $\Psi$ for the strongly localized gene interaction networks.  $\Psi$ takes $O(1)$ values only on the localized nodes, which consists of the hub node and its neighbors.  The complicated structures  of $\Psi$ are observed due to multiple heavily localized hub nodes of large degrees.  We also see several peaks around the center $\lambda =0$, which may be related to the eigenvalues of the star-graphs \cite{Bauer2001, Evangelou1992}.

\begin{figure}[!h]
\centering
\includegraphics[width=2.5in]{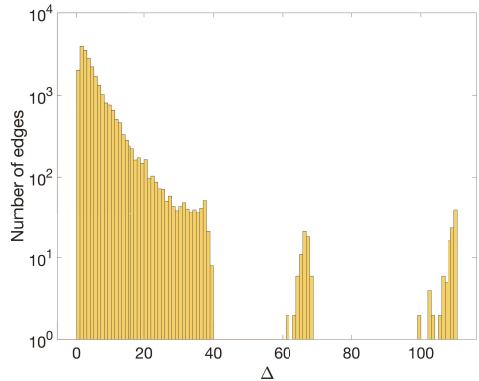}
\caption{Histogram of the degree discrepancies $\Delta$ between two nodes linked by an edge in ID:101 gene interaction network, which was inferred from the expression data from lung cancer cells. The bin width is $1$. The total number of edges is 24654.}
\label{Fig:ds}
\end{figure}

To study the topology of the gene interaction networks with multiple localized hubs of degree $d \gtrsim 40$, we show  a histogram of degree discrepancies $\Delta$ between two nodes on both ends of an edge in FIG.\ref{Fig:ds}.
\begin{equation}
\Delta = |k_i - k_j |,
\end{equation}
where $k_i$ and $k_j$ are the degrees of the connected $i$ and $j$ nodes, respectively.
Figure \ref{Fig:ds} is the result of the gene interaction network ID:101 obtained from  the gene expression experiments on cells from lung cancer.   Total number of edges is $24,654$ in this network and  there are two heavily localized nodes whose degrees are $d_1 =112$ and $d_2 =69$.  From FIG.\ref{Fig:ds} we see two distinct peaks near $\Delta \simeq 110$ and $70$, which should correspond to the edges in the sub-networks centered on the two hubs $d_1$ and $d_2$, respectively.  We find the naive histogram of $\Delta$ also shows the disassortative feature of the hub nodes, where $\Delta \simeq d$ for $\Delta \gtrsim 40$.  

\begin{figure}[h!]
\centering
\includegraphics[width=3.5in]{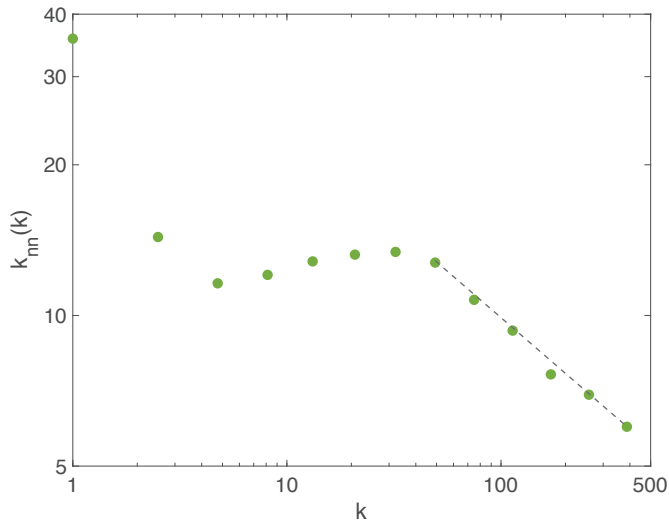}
\caption{The correlation function of the node degrees averaged over all edges in the 254 gene interaction networks.  The dashed line is the result of the fitting.  The negative power $k_{nn}(k)\propto k^{-0.37}$ describes the disassortative degree correlation exists in the sub-networks centered on the hub nodes whose degrees are $k \gtrsim 40$. }
\label{Fig:knn}
\end{figure}

In FIG.\ref{Fig:knn}, we plot the degree correlation function $k_{nn}(k)$ obtained from 254 gene interaction networks in log-log scale.  The averaging has been done for all edges in these networks.  We use the logarithmic bins again, where $n$-th bin edge is given by $k=(1.5^n-1)/0.5$.  In the large $k$ regime, we see the disassortative behavior which is apparent from the negative slope of $k_{nn}(k)$ for $k \gtrsim 40$. We also apply a curve fitting for the power $\eta$ of the slope, by which we get $k_{nn}(k)\propto k^{-0.37}$.  The fitting result is also shown in FIG.\ref{Fig:knn} by the dashed line.  In the medium $k$ regime where $6 \lesssim k \lesssim 40 $, we obtain the neutral or probably the assortative behavior of the correlation of the node degrees.  

From these observations,  we expect that the whole cancer gene interaction network contain multiple disassortative sub-networks centered on the hub whose degree is $d>40$. 

\begin{figure}[!h]
\centering
\includegraphics[width=3.5in]{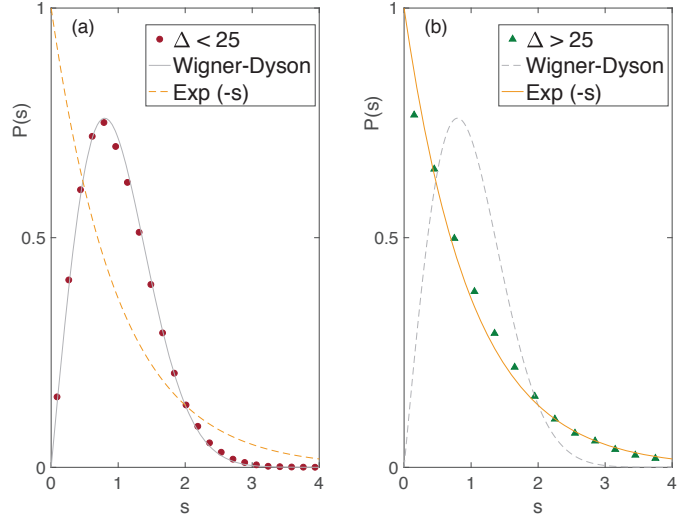}
\caption{{\rm (a)} The distribution of the nearest neighbor eigenvalues spacing $P(s)$ of the adjacency matrices of sub-networks which consists of edges that have $\Delta <25$ (the assortative or neutral degree correlation network).  $s$ is the distance between (unfolded) $\lambda_i$ and $\lambda_{i+1}$ normalized by the mean level spacing. The average has been taken over 27 networks listed in Table.\ref{Tab:01}. {\rm (b)} The $P(s)$ obtained from the sub-networks of edges $\Delta\ge 25$ (the disassortative degree correlation sub-networks localized on the hub nodes.)  }
\label{Fig:ps}
\end{figure}

The gene interaction networks which have a localized super-hub of degree larger than $d>125$ are selected for calculations of the distribution of nearest neighbor eigenvalues spacing $P(s)$. In each of such $27$ strongly localized networks with huge hubs, we extract edges according to the value of $\Delta$ (the degree discrepancies of linked nodes).  We divide each network into two sub-networks which consists of  edges with  (a) $\Delta < 25$ and (b) $\Delta \ge 25$, respectively.  FIG.\ref{Fig:ps}(a) shows $P(s)$ obtained from the sub-network (a).  The unfolding procedure has been done for each of 27 networks separately,  then we take the average of local $P(s)$.  The coincidence with the Wigner-Dyson distribution is very nice, which has been also tested by the one sample Kormogorov-Smirnov test.  We have tested $906$ segments of unfolded eignevalues over 27 networks and $93\%$ of them have p-values larger than the significance level $\alpha=0.05$.

We follow the same procedure for the sub-networks (b) which consist of the edges $\Delta\ge 25$, and show $P(s)$ in FIG.\ref{Fig:ps}(b).   The Poisson distribution of $P(s)$ is obtained in $76\%$ of total $502$ eigenvalue segments over 27 networks at the significance level $\alpha=0.05$, which also suggests the modular behavior of the strongly localized sub-networks.  We note that the number of edges which belong to the sub-network (b) is $4315$ ($31\%$ of total), for example in the  ID:191 gene interaction network. The number of contributing nodes are $3825$ ($48\%$ of total) in the same sub-network.  In Table \ref{Tab:01}, we show the statistics of the sub-networks (b) in each of 27 gene interaction networks.   The average ratio of edges consisting the hub centered communities $\Delta > 25$ is 19\%.

In NDEx \cite{Pratt2015}, we show a network graph of sub-network (b) of ID:230 gene interaction network in colon tumor. The URL accessible to the graph is given in Supplementary file.  It can be seen that hubs are connected via mediator nodes of relatively small degree, including the genes RECK, BCAS1, RNF168, etc.. A detailed study of the nature of these mediator genes is important for further understanding of the disease.

\section{Conclusions}
254 networks have been analyzed to obtain the average behavior of gene networks in human cancer cells. More than $9\times 10^6$ edges are included in the calculation of the degree distribution $P(k)$. The power-law behavior has been observed where $P(k) \propto k^{-3.8}$. 

The cancer gene networks are strongly localized on the hub gene when the degree of the hub $d_c > 26$ for the networks where the mean degree $\bar{k}=4.7$, and $d_c > 40$ in the networks of $\bar{k}=6.6$, respectively. The eigenvector centralities on the hub fit well on the theoretical line which was obtained for the model networks of random graphs with the hub.  However, the critical values of the hub degree $d_c$ become larger than the theoretical result $d_c = 2\bar{k}$.

The localized cancer gene interaction networks are the disassortative (hubs avoiding hubs) networks and the degree correlation function $k_{nn}(k)$ has a negative power in the region $k \gtrsim 40$.
This result is also consistent with the evaluation from the eigenvector centrality for the networks where $\bar{k}=6.6$.  

The localized hub communities in which the edges have large degree discrepancies $\Delta >25$  are extracted. They show the modular behavior which was also confirmed from the Poisson distribution of the nearest neighbor eigenvalues spacing $P(s)$.    The sub-network in which $\Delta < 25$ form a giant component (a hair ball) and we obtain the Wigner-Dyson distribution of $P(s)$ in this sub-network.

In the localized hub centered community, the hubs are linked by the mediator genes which have small degrees.  The investigation of the nature of these mediator genes might be helpful for the diagnosis and the decision of medical treatments for human cancers. 

\begin{table}[!h]
\caption{\label{Tab:01}
The statisitics of the localized sub-networks where $\Delta\ge 25$.  $\Delta$ is the difference of degrees of linked nodes.  }

 \begin{tabular}{lccccc}
\hline\hline
 Network  &	Total		& Edges of 	& Consisting  	& Ratio of 	& Ratio of   \\
    ID &	edges		&  $\Delta \ge 25$ 	& nodes	& edges	& nodes  \\
\hline
ID:191&	13787&		4315&	3825&	0.319&	0.478 \\
ID:248&	15961&		3478&	2786&	0.217&	0.348\\ 
ID:117&	19267&		1393&	1224&	0.072&	0.153\\ 
ID:247&	19270&		9254&	6652&	0.480&	0.831\\ 
ID:153&	19720&		12957&	6867&	0.657&	0.927\\ 
ID:243&	23300&		11456&	7073&	0.491&	0.885\\ 
ID:251&	26742&		2715&	2322&	0.101&	0.290\\ 
ID:111&	27525&		4014&	3608&	0.145&	0.451\\ 
ID:86&	28671&		4309&	3919&	0.150&	0.489\\ 
ID:148&	29215&		6535&	5237&	0.223&	0.654\\ 
ID:57&	29882&		4252&	3840&	0.142&	0.480\\ 
ID:143&	30714&		6879&	5210&	0.223&	0.651\\ 
ID:146&	31424&		5459&	4542&	0.173&	0.567\\ 
ID:125&	32213&		1900&	1767&	0.058&	0.220\\ 
ID:162&	35648&		8204&	5689&	0.230&	0.711\\ 
ID:203&	35965&		7589&	5450&	0.211&	0.681\\ 
ID:147&	37468&		5246&	4175&	0.140&	0.521\\ 
ID:109&	38016&		3804&	3282&	0.100&	0.410\\ 
ID:112&	39315&		2259&	2125&	0.057&	0.265\\ 
ID:119&	39754&		3594&	2988&	0.090&	0.373\\ 
ID:24&	43079&		4325&	3486&	0.100&	0.435\\ 
ID:210&	43184&		2965&	2623&	0.068&	0.327\\
ID:114&	44674&		5735&	4233&	0.128&	0.529\\
ID:179&	46849&		5943&	4246&	0.126&	0.530\\
ID:1&	49622&		4966&	3600&	0.100&	0.450\\
ID:221&	50594&		5896&	4217&	0.116&	0.527\\
ID:26&  51994&     4091&   3048&   0.078&  0.381\\ 
\hline\hline
\end{tabular}
\end{table}

\end{document}